\documentclass[12pt]{iopart}
\usepackage{graphicx}
\usepackage[english]{babel}
\usepackage{SIunits}
\usepackage{color}
\usepackage{epstopdf}
\usepackage{cite}

\begin{document}
\TDM

\title[Contact-induced charge contributions to non-local spin transport]{Contact-induced charge contributions to non-local spin transport measurements in Co/MgO/graphene devices}

\author{F. Volmer, M. Dr\"{o}geler, T. Pohlmann, G. G\"{u}ntherodt, C. Stampfer, B. Beschoten}
\address{2nd Institute of Physics and JARA-FIT, RWTH Aachen University, D-52074 Aachen, Germany}
\ead{bernd.beschoten@physik.rwth-aachen.de}

\begin{abstract}

Recently, it has been shown that oxide barriers in graphene-based non-local spin-valve structures can be the bottleneck for spin transport. The barriers may cause spin dephasing during or right after electrical spin injection which limit spin transport parameters such as the spin lifetime of the whole device. An important task is to evaluate the quality of the oxide barriers of both spin injection and detection contacts in a fabricated device. To address this issue, we discuss the influence of spatially inhomogeneous oxide barriers and especially conducting pinholes within the barrier on the background signal in non-local measurements of graphene/MgO/Co spin-valve devices. By both simulations and reference measurements on devices with non-ferromagnetic electrodes, we demonstrate that the background signal can be caused by inhomogeneous current flow through the oxide barriers. As a main result, we demonstrate the existence of charge accumulation next to the actual spin accumulation signal in non-local voltage measurements, which can be explained by a redistribution of charge carriers by a perpendicular magnetic field similar to the classical Hall effect. Furthermore, we present systematic studies on the phase of the low frequency non-local ac voltage signal which is measured in non-local spin measurements when applying ac lock-in techniques. This phase has so far widely been neglected in the analysis of non-local spin transport. We demonstrate that this phase is another hallmark of the homogeneity of the MgO spin injection and detection barriers.  We link backgate dependent changes of the phase to the interplay between the capacitance of the oxide barrier to the quantum capacitance of graphene.

\end{abstract}


\maketitle

\section{Introduction}
\begin{figure}[tb]
	\includegraphics{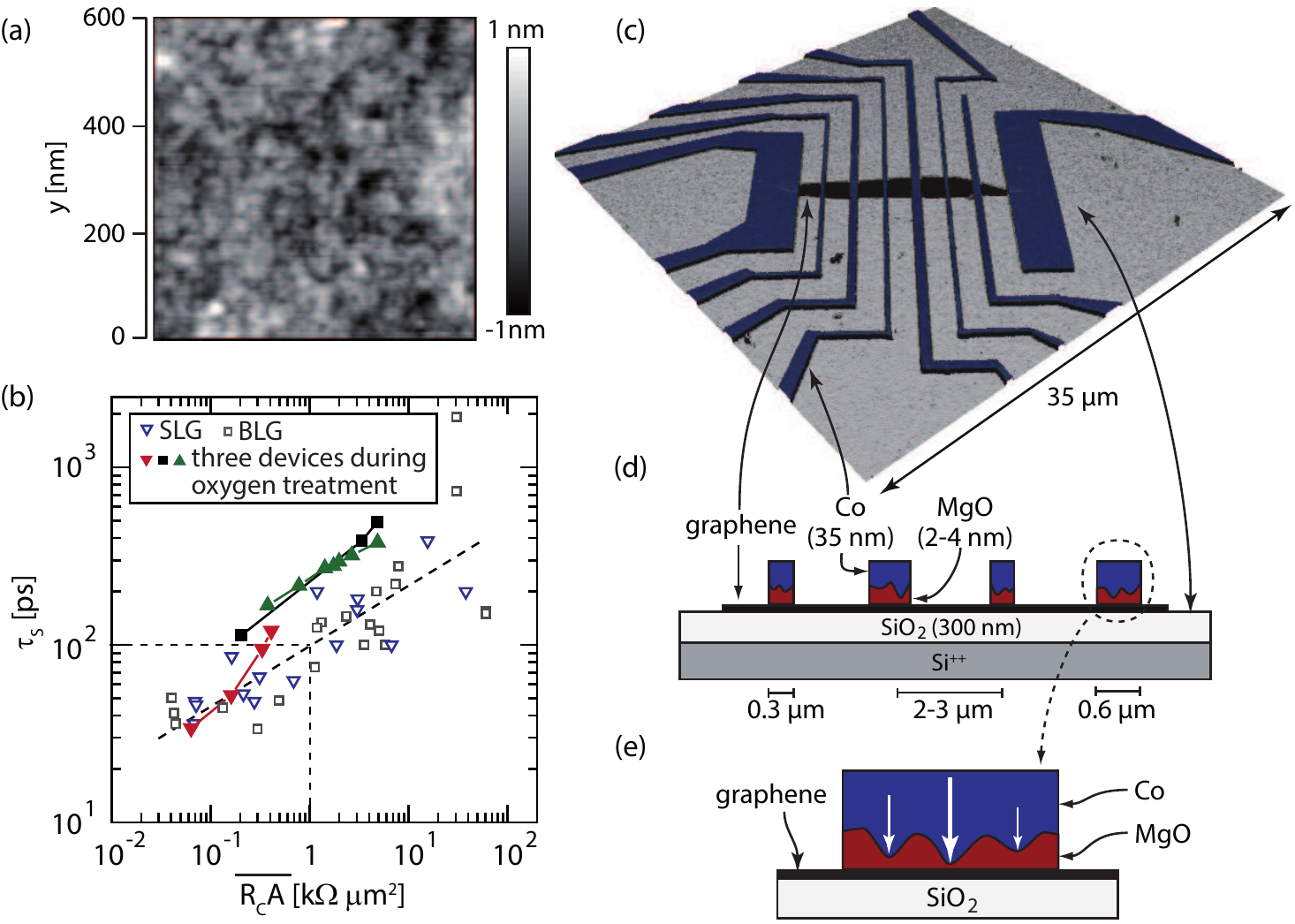}
	\caption{(Color online) (a) Atomic force microscopy image of \unit{3}{\nano\meter} MgO deposited on top of graphene showing peak-to-peak values of up to \unit{2}{\nano\meter}. (b) For our graphene/MgO/Co spin-valve structures the spin lifetime scales with the resistance area product of the contacts over a wide range for both single and bilayer graphene devices (data taken from \cite{PhysRevB.90.165403}). (c) False color atomic force microscope image of one of our spin-valve devices and (d) schematic cross section along the graphene flake demonstrating the general design. One important feature is the inhomogeneous oxide layer because of the Volmer-Weber island growth mode of MgO on graphene which favors pinholes for thin MgO layer thicknesses. (e) Schematic close-up of one contact illustrating the proposed spatially inhomogeneous current distribution (white arrows) over such barriers. }
	\label{fig:fig1}
\end{figure}

The oxide barrier between graphene and ferromagnetic spin injection and spin detection electrodes in non-local spin-valve structures can be a limiting factor for spin transport \cite{PhysRevLett.105.167202,PhysRevB.86.235408,Dlubak2012}. Recently, we conducted a systematic study of the role of the Co/MgO to graphene interface on the spin lifetime in both single-layer graphene (SLG) and bilayer graphene (BLG) in non-local geometry \cite{PhysRevB.88.161405,PhysRevB.90.165403}. We concluded that the island growth mode of a thin MgO layer grown directly onto graphene (see e.g. the atomic force microscopy image in figure \ref{fig:fig1}(a)) favors the formation of inhomogeneous oxide barriers and presumably pinholes, which in turn significantly influence both the charge transport in the local geometry and the spin transport in the non-local geometry. Furthermore, we demonstrated that an additional oxidation step can significantly improve the contact properties and with this the spin transport \cite{PhysRevB.90.165403}. In our studies we observed that the room temperature spin lifetimes ($\tau_s$) scale with the resistance area product of the spin injection and detection contacts ($R_cA$) over a wide range (figure \ref{fig:fig1}(b)), demonstrating that the contacts limit the spin lifetime in not perfect oxide barriers. This result leads to the question of how to evaluate the quality of the oxide barrier in a fabricated device next to the so-called Rowell criteria for tunnel barriers, which are the exponential increase of resistance with oxide thickness, a non-linear $I$-$V$-curve, and a decreasing conductance at lower temperatures \cite{Akerman2002}. Additional criteria might be helpful as the Rowell criteria were questioned in several publications\cite{jonsson-akerman:1870,Rabson2001,Oliver2003}.
Amongst others, it was discussed that these criteria can still be fulfilled although the oxide barriers under investigation exhibit pinholes. A different approach by Han et al. \cite{PhysRevLett.105.167202} evaluates the quality of the oxide barriers by linking contact properties (transparent contacts, tunneling contacts, or intermediate contacts with pinholes) to qualitatively different charge carrier density dependence of the non-local spin signal. However, in one of our previous studies \cite{Droegeler2014} we found hints that the charge carrier density dependence of the spin signal alone is not an unambiguous sign for tunneling behavior of the contacts.

Here, we show that inhomogeneous oxide barriers nevertheless still leave distinct fingerprints in non-local spin measurements such as charge accumulation underneath the spin detection electrode. A non-local four-terminal geometry is often chosen over two- or three-terminal geometries as it is believed to probe pure spin voltage avoiding any charge effects. But previous publications already demonstrated that e.g. the interplay of Peltier and Seebeck effects \cite{PhysRevLett.105.136601} or spatially inhomogeneous injection and detection of charge carriers over the electrodes \cite{PhysRevB.76.153107} can result in a non-local baseline resistance in spin-valve measurements. We pick up the latter work and  expand it to the case of Hanle spin precession measurements by including an additional perpendicular magnetic field. With this we are able to simulate the non-linear background signal in the Hanle spin precession measurements (see e.g. figures \ref{fig:fig2} (a)-(c)), which is also observed in many other publications \cite{PhysRevLett.101.046601,doi:10.1021/nl301050a,PhysRevB.80.214427,PhysRevB.87.081402,PhysRevB.87.075455,PhysRevLett.104.187201,han222109,PhysRevLett.109.186604,PhysRevLett.113.086602}. We will show that this background signal can be attributed to a redistribution of charge carriers in the perpendicular magnetic field similar to the classical Hall effect. In some publications such background is also attributed to the fact that with increasing perpendicular magnetic field the magnetization of the ferromagnetic electrodes rotates out-of-plane \cite{PhysRevLett.101.046601,Idzuchi2012,PhysRevLett.113.086602}. Although this effect might also be relevant at larger magnetic fields, we want to emphasize that this magnetization rotation can only account for spin signals symmetric in magnetic field. In contrast, we usually observe an antisymmetric background signal which is linear in $B$ in addition to a symmetric signal which varies quadratically with magnetic field. Both contributions can be explained with our model based on charge accumulation.

In the second part of this paper we discuss another interesting effect which may also be utilized to classify the quality of the oxide barrier. For our charge and spin transport measurements we use ac lock-in techniques. We observe that the lock-in phase of the non-local voltage signal may strongly deviate from the phase of the actual spin signal, which can be explained by a phase-shifted charge signal. The phase and magnitude of this charge signal seem to be highly related to the contact properties of the graphene/MgO/Co interface and therefore may be another fingerprint indicating the quality of the oxide barriers. Although the physical origin of the phase is still under investigation, we will show a possible link to the interplay of oxide barrier capacitance and the quantum capacitance of graphene.

\section{Device fabrication and measurement techniques}
\label{fabrication}

All devices were fabricated from exfoliated SLG and BLG. After the deposition of the flakes onto $\textrm{SiO}_2$(\unit{300}{\nano\meter})/$\textrm{Si}^{++}$ wafers they are put into acetone and thereafter into isopropyl alcohol to remove possible organic contaminations. As a next step the e-beam lithography is carried out with PMMA dissolved in ethyl lactate and n-butyl acetate. The developer is a mixture of isopropyl alcohol and methyl isobutyl ketone. Prior to the deposition of the electrodes, the samples are stored in ultra high vacuum for several days to allow an outgassing of the above-mentioned chemicals and water residuals. We use electron-beam evaporation from MgO crystals (99.95\% metals basis) and Co pellets (99.95\% metals basis) at a base pressure of $\unit{1 \times \power{10}{-10}}{\milli\bbar}$. The deposition rates are $\unit{0.005}{\nano\meter\per\second}$ and $\unit{0.015}{\nano\meter\per\second}$ for MgO and Co, respectively, both at an acceleration voltage of \unit{4.5}{\kilo\volt}. We first grow the MgO barrier with varying thicknesses up to 3~nm followed by 35-nm-thick ferromagnetic Co electrodes. The layout of our devices can be seen in figures \ref{fig:fig1}(c) and (d). An atomic force microscopy image of \unit{3}{\nano\meter} MgO grown on top of graphene is depicted in figure \ref{fig:fig1}(a) and demonstrates the island growth mechanism. Such kind of oxide layer leads to spatially varying injection of the spin-polarized current over the barrier (see white arrows in \ref{fig:fig1}(e)), which becomes important for the explanation of the non-local magnetic field dependent background signal in section \ref{background}.

Hanle spin precession measurements are performed in non-local 4-terminal geometry (figure \ref{fig:fig2}(d)) with the external magnetic field $B$ applied perpendicularly to the graphene sheet. All measurements are performed at room temperature under vacuum conditions at a base pressure of $\unit{6 \times \power{10}{-4}}{\milli\bbar}$. We use standard ac lock-in techniques, where the reference signal of the lock-in modulates the current through the device at a frequency of $\unit{18}{\hertz}$ with rms-values of up to $\unit{20}{\micro\ampere}$ (more information about the measurement setup in section \ref{phase}).

Selected Hanle curves are depicted in figures \ref{fig:fig2}(a)-(c) for device A (see next section) with both parallel and anti-parallel alignments of the inner Co electrodes. The Hanle depolarization curves can be described by the steady-state Bloch-Torrey equation \cite{ISI000249789600001,PhysRevB.37.5312}:
\begin{equation}
	\frac{\partial \vec{s}}{\partial t}\;=\;\vec{s}\times \vec{\omega}_0+D_{\text{s}}\nabla^2\vec{s}-\frac{\vec{s}}{\tau_{\text{s}}}\;=0,
\end{equation}
where $\vec{s}$ is the net spin vector, $\vec{\omega_0}=g\mu_{B} \vec{B}/\hbar$ is the Larmor frequency, where we assume $g=2$, $D_{\text{s}}$ is the spin diffusion constant, and $\tau_{\text{s}}$ is the transverse spin lifetime. With $L$ being the distance between spin injection and spin detection electrodes, we define the following dimensionless parameters: $b\equiv g \mu_{\text{B}} B \tau_{\text{s}}/ \hbar$, $l\equiv L/\sqrt{2 D_{\text{s}} \tau_{\text{s}}}$ and $f(b)=\sqrt{1+b^2}$. With a simplified analytical solution we use the following fit function to describe the Hanle curves and to extract the spin lifetimes \cite{ISI000249789600001,PhysRevB.37.5312}:

\begin{equation}
\fl
\begin{array}{cl}
R_{\textrm{nl}}^{\textrm{Hanle}} = & \Delta R_{\textrm{nl}}\; \frac{1}{2f(b)} \left[ \sqrt{1+f(b)}\; \text{cos}\left( \frac{l b}{\sqrt{1+f(b)}} \right)- \frac{b}{\sqrt{1+f(b)}}\; \text{sin}\left( \frac{l b}{\sqrt{1+f(b)}} \right)  \right] \\ & \times\text{exp}\left( -l \sqrt{1+f(b)} \right).
\end{array}
\end{equation}

In addition to the pure Hanle spin precession signal, there is the above-mentioned $B$ dependent background signal which we observe in most measurements (see figures \ref{fig:fig2}(a)-(c)). This background can be well fitted by a polynomial of second order with coefficients $c_i$ (red curves in figures \ref{fig:fig2}(a) and (b)). Its origin will be discussed in the next section. The total non-local resistance thus equals to:

\begin{equation}
\label{backgroundfit}
    R_{\textrm{nl}}^{\textrm{total}}(B) \;=\; R_{\textrm{nl}}^{\textrm{Hanle}}(B) + c_2 B^2 + c_1 B + c_0 \;.
\end{equation}

\section{Magnetic field dependent background in Hanle spin precession measurements}
\label{background}

\begin{figure}[tb]
	\includegraphics{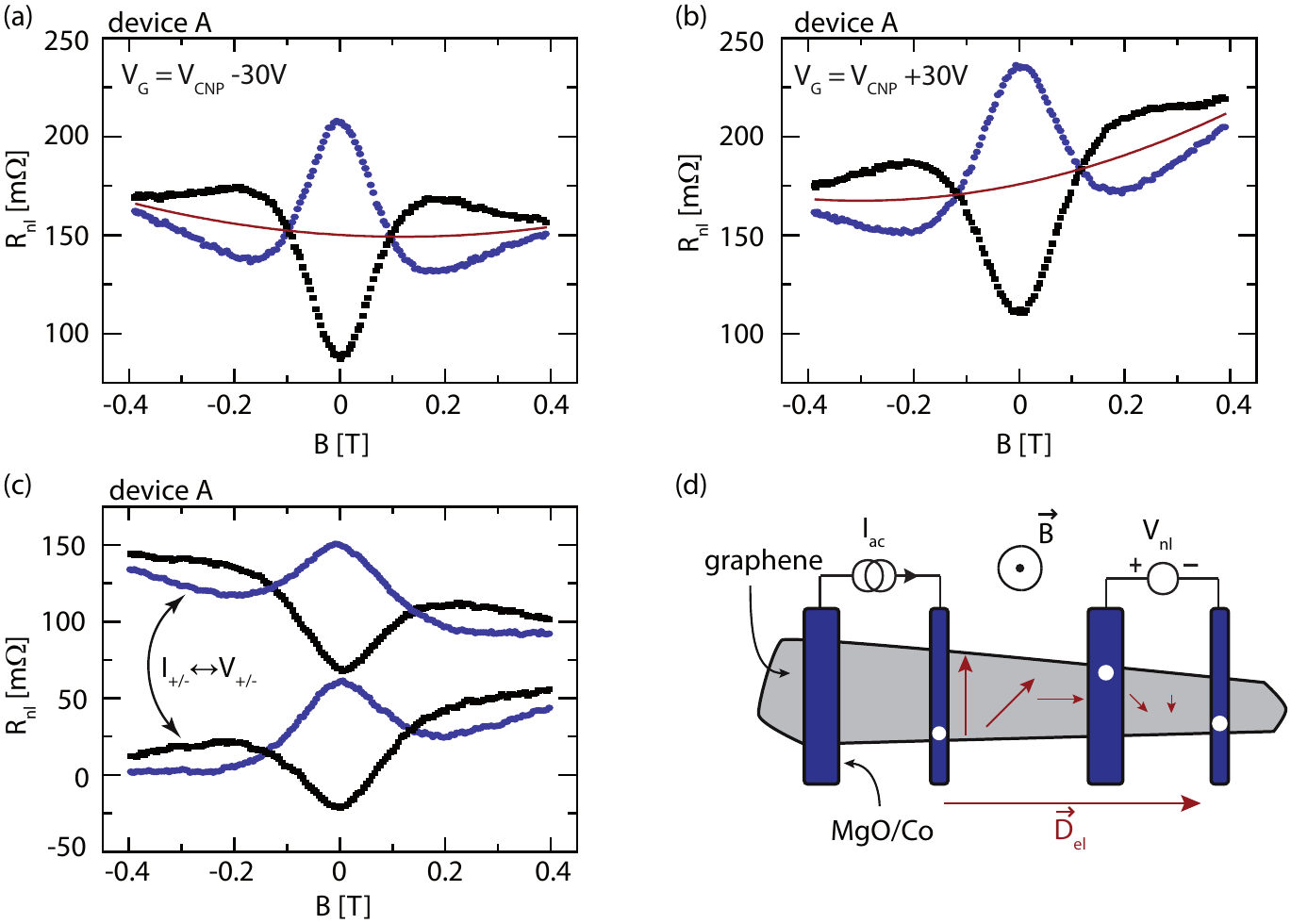}
	\caption{(Color online) (a) and (b): Room temperature Hanle spin precession curves of device A measured at $V_{\text{G}}=V_{\text{CNP}}\pm \unit{30}{\volt}$ which gives equal carrier densities in the hole (a) and electron regime (b), respectively. A sign reversal in the linear component of the non-local background (red curve) is clearly observed. (c) A similar behavior is seen if the injection and detection circuits is swapped. These curves are plotted as measured. No vertical offset was added. This implies that the linear component of the background is sensitive to both to the type of carriers and the direction of carrier diffusion with respect to the magnetic field. The latter is depicted as $\vec{D}_{el}$ in the schematic device layout and wiring for the non-local Hanle precession measurement in (d). White circles illustrate possible positions of the conducting pinholes in the MgO barrier near the edges of the graphene flake, resulting in a measurement geometry that is sensitive to a non-local Hall effect. $\vec{D}_{el}$ is the direction of diffusion. }
	\label{fig:fig2}
\end{figure}

We first discuss experimental results that are crucial to understand the charge-driven origin of the $B$ dependent background in Hanle spin precession curves. In particular the linear term ($c_1$ in equation 3) of the background strongly depends on the applied gate voltage or, equivalently, the charge carrier density $n$, which is calculated by $n=\alpha \left(V_{\text{G}} -V_{\text{CNP}}\right)$ according to the capacitor model with $\alpha \approx \unit{7.18 \times \power{10}{10}}{\reciprocal\volt\centi\meter\rpsquared}$ and $V_{\text{CNP}}$ being the gate voltage at the charge neutrality point (CNP). Typically, we observe that the linear term switches sign when going from the electron to the hole regime. Corresponding Hanle curves for  are shown in figures \ref{fig:fig2}(a) and (b) for device A for hole ($V_\text{G}=V_{\text{CNP}}-\unit{30}{\volt}$) and electron ($V_\text{G}=V_{\text{CNP}}+\unit{30}{\volt}$) doping, respectively. All other measurement parameters were kept identical for both measurements. The change in the sign of the linear component $c_1$ is the first important hint that this part of the background is due to charge and does not result from spin effects.

Furthermore, there is even a sign reversal in the linear term if the injection and detection circuits are interchanged, i.e. the injection electrodes $I_+$ and $I_-$ become the detection electrodes $V_+$ and $V_-$ and vice versa (figure \ref{fig:fig2}(c)). The injected spin polarized charge carriers diffuse through the graphene sheet from the injection to the detection electrodes and by interchanging injection and detection contacts the direction of diffusion ($\vec{D}_{el}$ in figure \ref{fig:fig2}(d)) reverses into the opposite direction, i.e. $-\vec{D}_{el}$. Accordingly, the non-local background depends on (1) the type of charge carriers (electrons or holes) and (2) the motion of the carriers relative to the perpendicular magnetic field direction. This is typical for the classical Hall effect which results in a charge accumulation transverse to the drift and diffusion direction. We note that our devices do not exhibit any Hall geometry at first glance as all of the electrodes cover the whole width of the graphene flake which is expected to shorten any transverse voltage from Hall effect. In contrast, a transverse Hall voltage might be detected in the non-local voltage signal if conducting pinholes within the highly resistive oxide barriers were placed near opposite edges of the graphene channel as indicated by the white circles in figure \ref{fig:fig2}(d). In agreement with this assumption, the amplitude of the linear background is less pronounced (not shown) in devices with larger contact resistance area products, which exhibit less pinholes and yield longer spin lifetimes (see our previous studies in references \cite{PhysRevB.88.161405,PhysRevB.90.165403}).

\begin{figure}[tb]
	\includegraphics{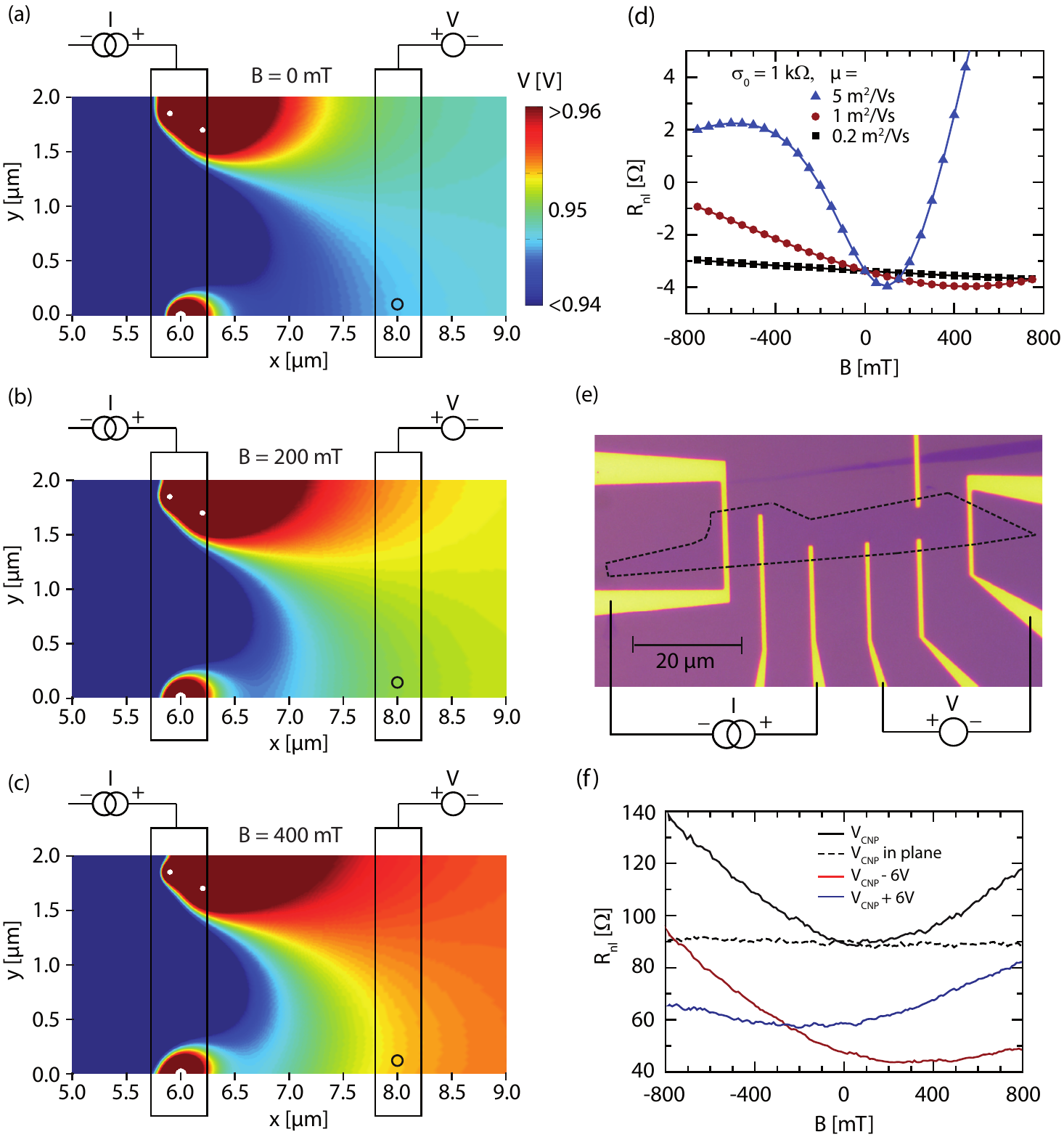}
	\caption{(Color online) (a)-(c) The non-local potential distribution simulated for a device geometry which is used in our actual spin precession measurements at different magnetic fields. The color bar in (a) is also valid for (b) and (c) and its range was adjusted in such a way that the potential landscape in the non-local part of the device is visible (values higher than \unit{0.96}{\volt} or lower than \unit{0.94}{\volt} were cut).  The white circles at the injection electrode $I_+$ represent pinholes which are put to a potential of \unit{1}{\volt}. For simplicity it was assumed that $V_+$ probes the voltage over only one pinhole, whose position is depicted as the black circle. (d) The simulated non-local background signal for this geometry for different values of the charge carrier mobility.  (e) Optical image of a graphene flake (edge highlighted by dashed line) with Cr/Au-contacts. The depicted measurement geometry yields the non-local resistances in (f) for in- and out-of-plane magnetic fields at different charge carrier densities. }
	\label{fig:fig3}
\end{figure}

To support this notion, we simulate the non-local measurements by applying the model of inhomogeneous injection and detection of electrons proposed by Johnson and Silsbee in reference \cite{PhysRevB.76.153107}. We assume that the whole injected current at the $I_+$ electrode flows over conducting pinholes (white circles in figure \ref{fig:fig3}(a)) and put all pinholes of $I_+$ to the same potential of  \unit{1}{\volt} while the electrode $I_-$ is set to \unit{0}{\volt}. The electrodes $I_-$ and $V_-$ are placed at $x= \unit{0}{\micro\meter}$  and $x= \unit{15}{\micro\meter}$, respectively. Both are not shown in figure \ref{fig:fig3}(a) as we visualize the interesting regime near $I_+$ and $V_+$.  Far away from the injection electrode $I_+$ the current density will flow uniformly towards $I_-$ in $-x$-direction. However, nearby the $I_+$ electrode the current flows in a complicated manner radially away from the pinholes. Consequently, there are also charge currents on a microscopic scale even in the graphene part between electrodes $I_+$ and $V_+$. Of course, the net current along the x-axis in this part of the device has to be zero under steady state condition. Nevertheless the local currents generate a spatially varying charge potential.

Although the changes in this potential landscape rapidly decrease in the direction of the detection electrodes, they can even reach the non-local voltage probe, i.e. the $V_+$ electrode, which is \unit{2}{\micro\meter} away (see e.g. figure \ref{fig:fig3}(a)). For simplicity, we assume that this detection electrode only probes the voltage at one point which acts as a detection pinhole (black open circle in figure \ref{fig:fig3}(a)). If we assume more than one pinhole in the $V_+$ electrode, they will all be short-circuited by the metallic toplayer. Hence, a current will flow to compensate possible potential differences in the graphene sheet underneath these pinholes, which makes the simulation far more complicated. With this model, Johnson and Silsbee \cite{PhysRevB.76.153107} are able to explain the so-called baseline resistance in non-local spin-valve measurements, i.e. an offset in the spin signal like it is seen and discussed in figure \ref{fig:fig6}(a).

To account for the charge-induced background in Hanle spin precession measurements, we now extend the model and include the effect of a perpendicular magnetic field on the charge accumulation in the non-local detection circuit. The Lorentz force will affect the charge currents and therefore modify the potential landscape in the non-local voltage probing part of the device. The results are shown for two magnetic fields in figures \ref{fig:fig3}(b) and (c) for the same device geometry as in figure \ref{fig:fig3}(a). For the simulation we numerically solved the steady state potential equation:
\begin{equation}
\label{model}
\nabla \left( \sigma \nabla \Phi \right) = 0, \ \textrm{with} \ \sigma = \frac{\sigma_0}{1+\left( \mu B \right)^2}
\left[
\begin{array}{cc}
  1 & \mu B \\
  -\mu B & 1
\end{array}
\right]
\end{equation}
by means of a finite element method. Here, $\Phi$ is the electric potential, $\sigma$ is the $B$ field dependent conductivity tensor which is used to describe the Hall effect with $\sigma_0$ being the conductivity at $B=0$~T and $\mu$ being the charge carrier mobility.

The non-local voltage between $V_+$ and $V_-$ is normalized by the total current density $j=\sigma \nabla \Phi$ which flows in the injection circuit to get the non-local resistance $R_{\text{nl}}$. In figure \ref{fig:fig3}(d) we show simulated magnetic field dependent $R_{\text{nl}}$ curves for the device geometry of figure \ref{fig:fig3}(a) for different charge carrier mobilities assuming a constant sheet resistance of graphene of $R_\text{sq}=1/\sigma_0=1/(n e \mu)=\unit{1}{\kilo\ohm}$. The overall magnitude of several $\Omega$ is in good agreement to experimental data. The magnetic field dependence can well be described by a second order polynominal function for lower carrier mobilities (black and red curves in figure \ref{fig:fig3}(d)), which is in qualitative agreement with the experimental data in figure \ref{fig:fig2}. However, at larger carrier mobilities, there are deviations from this simple dependence at larger magnetic fields (blue curve in figure \ref{fig:fig3}(d)).  Furthermore, the minima shift towards $B=\unit{0}{\tesla}$ with increasing mobilities. In fact, the mobility just amplifies the effect of the magnetic field (see the terms $\mu B$ in equation \ref{model}) and therefore the minimum of the curve which is at \unit{500}{\milli\tesla} for \unit{1}{\square\meter\per\volt\second} is shifted to \unit{100}{\milli\tesla} for \unit{5}{\square\meter\per\volt\second}. Of course there will be an upper limit to $\mu B$ at which our model breaks down, as equation \ref{model} is only valid for diffusive Boltzmann transport and will thus not apply for devices showing charge transport near the ballistic regime (i.e. devices with high charge carrier mobilities at low temperatures). Furthermore, other magneto-resistive effects have to be considered at large magnetic fields.

Hence, it is not clear how reliable the decrease in the non-local resistance is at large negative magnetic fields for the high mobility sample. Nevertheless, we note that usually the magnetic field range is not more than $\unit{\pm 400}{\milli\tesla}$ in Hanle spin precession measurements (see e.g. figure \ref{fig:fig2}). This holds for devices with spin lifetimes down to less than 100 ps. For our high mobility devices which exhibit nanosecond spin lifetimes \cite{Droegeler2014} there is even a much smaller magnetic field range needed ($\unit{\pm 200}{\milli\tesla}$). Within this field range a fit of the non-local background with a polynomial of second order like in equation \ref{backgroundfit} is again a good approximation.

It is important to note that the $B$ field dependence and magnitude of the non-local contribution by magnetic field-driven charge accumulation is rather sensitive to the shape and spatial distribution of the pinholes. For example, an accumulation of the pinholes near the edges of the graphene flake typically yields non-linear background signals like the ones in figures \ref{fig:fig2}. On the other hand, a more homogeneous distribution of the pinholes over the whole width of the graphene flake results in a more linear background. In general, a more inhomogeneous current injection (less and more randomly distributed pinholes) leads to a larger magnitude of the background. This is also in agreement with the above mentioned experimental result that the background is less pronounced in devices with higher quality oxide barriers showing long spin lifetimes.

To back-up our claim that one part of the observed background in non-local measurements does not result from spin transport but can be explained by the interplay of inhomogeneous charge injection and the redistribution of the currents by a perpendicular magnetic field, we fabricated a device with non-magnetic electrodes (figure \ref{fig:fig3}(e)). For this device we did not use any MgO barrier and deposited Cr/Au directly onto the graphene sheet.

To achieve inhomogeneous current injection and detection we fabricated electrodes which end on the graphene flake. With the wiring depicted in figure \ref{fig:fig3}(e), we measured the non-local resistance (non-local voltage normalized by the applied current in the injection circuit) as shown in figure \ref{fig:fig3}(f). When applying an in-plane magnetic field, which represents the situation of a spin-valve measurement, we observe a constant baseline resistance consistent to reference \cite{PhysRevB.76.153107} (see dashed line in \ref{fig:fig3}(f)). On the other hand, an out-of-plane field, which is used in Hanle spin precession measurements, leads to a distinctive magnetic field dependent non-local resistance, which can again well be described by a polynominal of second order. The quadratic contribution is most pronounced near the CNP. Most strikingly, the sign of the linear contribution $c_1$ again reverses between electron (blue curve) and hole (red curve) doping, which is in accordance to figures \ref{fig:fig2}(a) and (b). Also the sign reversal upon switching injection and detection circuits as shown in figure \ref{fig:fig2}(c) for the non-local spin-valves could be reproduced (not shown). We note that our simulation in general predicts a larger non-local signal with increasing width of the graphene flake as long as the injection and detection occurs close to the edges. Therefore, the large non-local resistance in figure \ref{fig:fig3}(f), which is much larger than what we usually observe in spin precession measurements, only results from the large graphene flake which we used in this experiment.

The design of the reference device was chosen to probe the validity of our model as for spin-valve structures with oxide barriers we have no information about the number and distribution of pinholes. As the Cr/Au contact area is equipotential, we can model the non-uniform current distribution more straightforwardly with the geometrical constrains of the electrodes. Both the mobility and the conductivity needed for the simulation were measured in four-terminal geometry. We are able to reproduce the experimental data in figure \ref{fig:fig3}(f) in a qualitative but not quantitative way. Deviations may stem from other charge effects like the above-mentioned interplay of Peltier and Seebeck effects \cite{PhysRevLett.105.136601}. But more importantly, in our model we only assume a single conductivity, one mobility, and in particular only one type of charge carriers within the graphene layer. Therefore, it is not surprising that our model fails to describe the non-local resistance near the CNP. In the device of figure \ref{fig:fig3}(e), we observe spatially varying $p$-doping in different regions of the flake. Right after transferring the contacted device into vacuum the CNPs of the different regions were at back gate voltages ranging between 25 and 30~V. After one day in vacuum there had been sufficient outgassing of the device (most likely of oxygen and water) and the new CNPs moved to gate voltages between $\unit{0-16}{\volt}$. As there is still significant doping in certain parts of the device, there are remaining spatially varying contaminations (most likely resist residues) which result in an inhomogeneous potential landscape. As a result, there will be patches of both electron and hole doped graphene parts near the CNP. However, such patches with different carrier densities and carrier type with presumably different mobilities are far beyond the scope of our model.

Furthermore, the resist residues between graphene and the Cr/Au-electrodes because of the e-beam lithography \cite{doi:10.1021/nl203733r,doi:10.1021/jz201653g} may also lead to an inhomogeneous charge injection and distribution into graphene underneath the electrodes. For simplicity, we assumed a constant potential of the graphene underneath the metal in our simulations. We additionally note that $p$-$n$-junctions, which are formed near the edges of the electrodes and are a result from doping by the metal \cite{PhysRevB.79.245430}, are not included in our model and may be a reason for the quantitative deviations between experiment and simulation. Nevertheless, both the simulation and the back-up experiment with non-ferromagnetic electrodes can qualitatively reproduce the non-local background in Hanle spin precession measurements from graphene/MgO/Co spin-valve devices, including the change in background signal for different charge carrier regimes and the pronounced parabolic dependence near the CNP.

\section{Phase between spin and charge signal in non-local geometry}
\label{phase}

In this section we discuss the phase difference which arise between the spin and charge voltage measured in non-local spin-valve geometry using an ac lock-in technique. We show that the phase and the magnitude of the charge voltage is connected to the electronic properties of the Co/MgO-to-graphene interface which therefore provides another fingerprint to characterize the quality of the oxide barriers. We furthermore demonstrate that the magnitude of the phase seems to be determined by both the oxide barrier capacitance and the quantum capacitance of graphene.

\begin{figure}[tb]
	\includegraphics{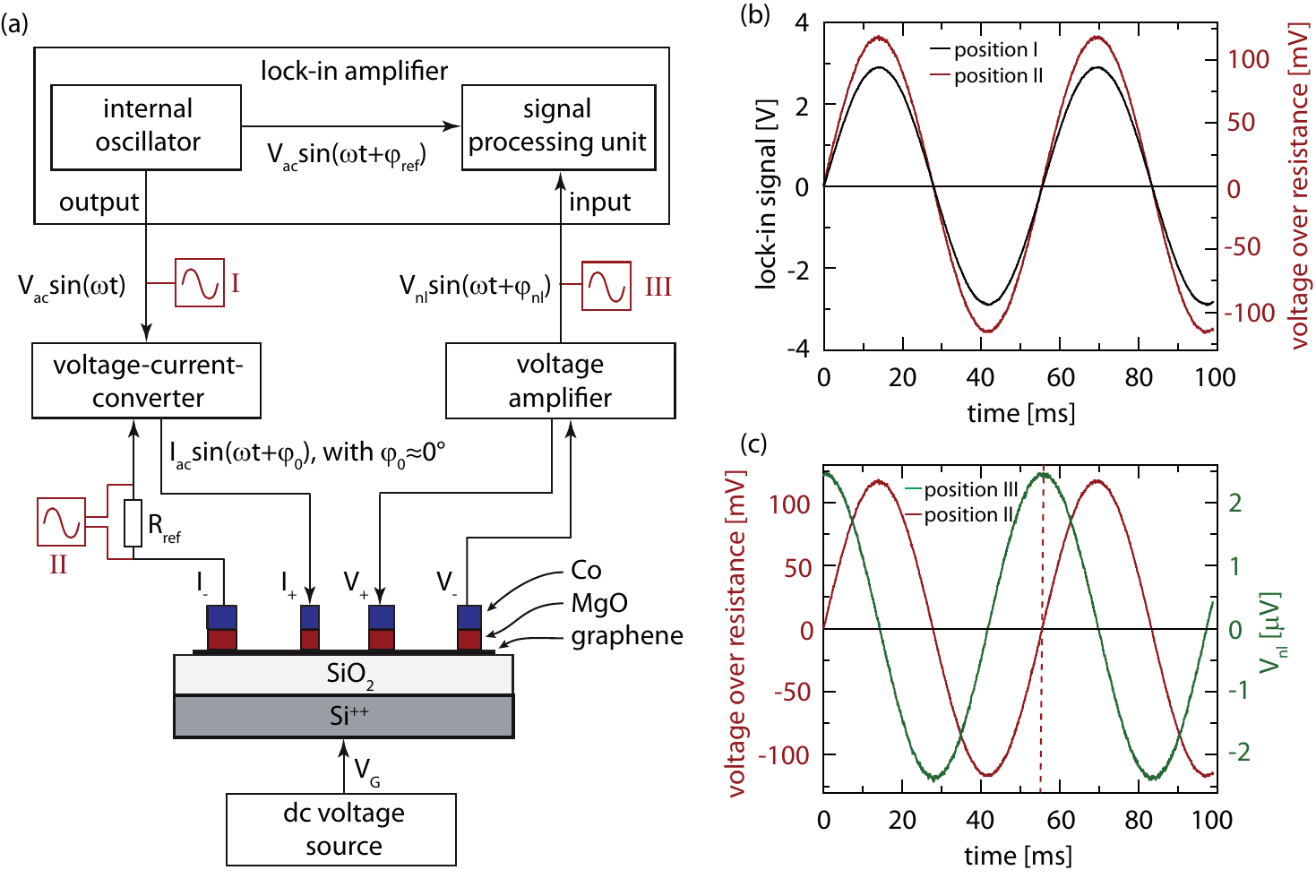}
	\caption{(Color online) (a) Schematic diagram of our setup defining the phases and signals at different positions in our circuit. The voltage-current-converter converts the voltage signal of the lock-in output into a current signal with adjustable amplitude, which is in phase to the lock-in signal for resistances and capacitances typical for our devices. In contrast, the non-local voltage measured between $V_+$ and $V_-$ usually exhibit a phase shift $\varphi_{\text{nl}}$ relative to position I. (b) and (c): Signals measured with an oscilloscope at three different positions which are depicted in (a). The signals stem from a spin-valve device with low ohmic, transparent contacts (flat d$V$-d$I$-curve, $R_cA<\unit{1}{\kilo\ohm\micro\meter\squared}$) and corresponding low spin lifetimes $\tau_s$ which vary between 70 and 100~ps. }
	\label{fig:fig4}
\end{figure}

In order to define the phase, we first discuss our setup (figure \ref{fig:fig4}(a)). We use the sinusoidal voltage signal $V_{\text{ac}} \textrm{sin}(\omega t)$ of the internal oscillator from a dual phase lock-in amplifier to drive a voltage-current-converter. The current $I_{\text{ac}} \textrm{sin}(\omega t+\varphi_0)$ is applied over the injection electrodes $I_+$ and $I_-$. The lock-in voltage output is measured at point I and is shown as the black curve in figure \ref{fig:fig4}(b). We note that in all local measurements such as gate-dependent graphene resistance traces or d$V$-d$I$-curves of the contacts (exact wiring for these measurements are given in reference \cite{PhysRevB.90.165403}) the relative phase between current and probing voltage does nor differ by more than \unit{3}{\degree}.  As long as we limit our measurements to low frequencies (we use $f=\unit{18}{\hertz}$) this is even valid for devices with large contact resistances ($R_cA>\unit{1}{\kilo\ohm\micro\meter\squared}$) with pronounced non-linear d$V$-d$I$-curves indicating tunneling behavior. This finding is illustrated by the time-dependent local voltage drop as measured by an oscilloscope over an additional reference resistance $R_{\text{ref}}$ which is in series with the device in the injection circuit at point II. The current over this resistance (red curve) and thus over the device is therefore in phase ($\varphi_0 \approx 0$) with the lock-in voltage output (black curve).

This can be understood from a simple theoretical consideration: The simplest equivalent circuit of an oxide barrier is a capacitor $C$ and a resistance $R$ connected in parallel. We approximate the capacitance of the MgO oxide barrier by the plate capacitor model with the permittivity of bulk MgO. This results in a capacitance of approximately \unit{\power{10}{-1}-\power{10}{-2}}{\farad\per\square\meter} with oxide layer thicknesses between \unit{1-3}{\nano\meter}. Calculating the impedance of the equivalent circuit leads to a phase between voltage and current over such a barrier of $\varphi=\textrm{arctan}(2\pi fRC)$. For typical contact areas in the order of some \unit{}{\micro\meter\squared}, contact resistances of up to \unit{100}{\kilo\ohm}, and a frequency of \unit{18}{\hertz} we obtain phase differences of under \unit{1}{\degree}. The larger phase of \unit{3}{\degree} in our experiment may result from additional capacitances in our setup.

For the non-local measurements the voltage difference between electrodes $V_+$ and $V_-$ is first amplified and then sent to the input channel of the lock-in. Its time-dependence can be measured by an oscilloscope at position III in our setup. As an example, we depict the green curve in figure \ref{fig:fig4}(c), which was taken on the same device as for the local measurements in figure \ref{fig:fig4}(b). This non-local voltage signal shows a pronounced phase difference of approximately $\unit{90}{\degree}$ relative to the current input signal. On the other hand, the same device exhibits the above-mentioned small phase difference in local measurements of only $\unit{3}{\degree}$. This implies that the phase $\varphi_{\text{nl}}$ in non-local measurements is not obviously connected to the capacitance and resistance of the graphene/MgO/Co contact.

As described in section 3, the non-local voltage is comprised of both charge- and spin-driven signals. In the following, we will demonstrate that only the charge contribution in the non-local voltage can exhibit large phase shifts while the spin signal (as later shown in figure \ref{fig:fig6}) is in phase with the injection current. For the latter, this is not surprising as the spin signal is a measure to the spin accumulation underneath the detection electrode, which in turn is built up by the spin polarized current over the oxide barrier. Because of the low measurement frequency, the time that the spins need to diffuse from the injection to the detection electrode can be ignored as the diffusion time cannot be much longer than the shortest measured spin lifetime, which is in the order of several ten ps. On the other hand, such a diffusion time is several orders of magnitude shorter than the ac current period and therefore cannot yield a considerable phase.

The lock-in amplifier allows to separate charge and spin driven contributions in the following way: At the signal input channel of the lock-in both the amplitude $V_{\text{nl}}$ and phase $\varphi_{\text{nl}}$ of the non-local signal is analyzed with respect to the reference signal by the signal processing unit. In basic operation, the sine wave of the internal oscillator is used as the reference signal. For the data analysis it is, however, important that this reference signal can be tuned by an additional phase $\varphi_{\text{ref}}$ relative to the phase of the internal oscillator (see figure \ref{fig:fig4}(a)).

\begin{figure}[tb]
	\includegraphics{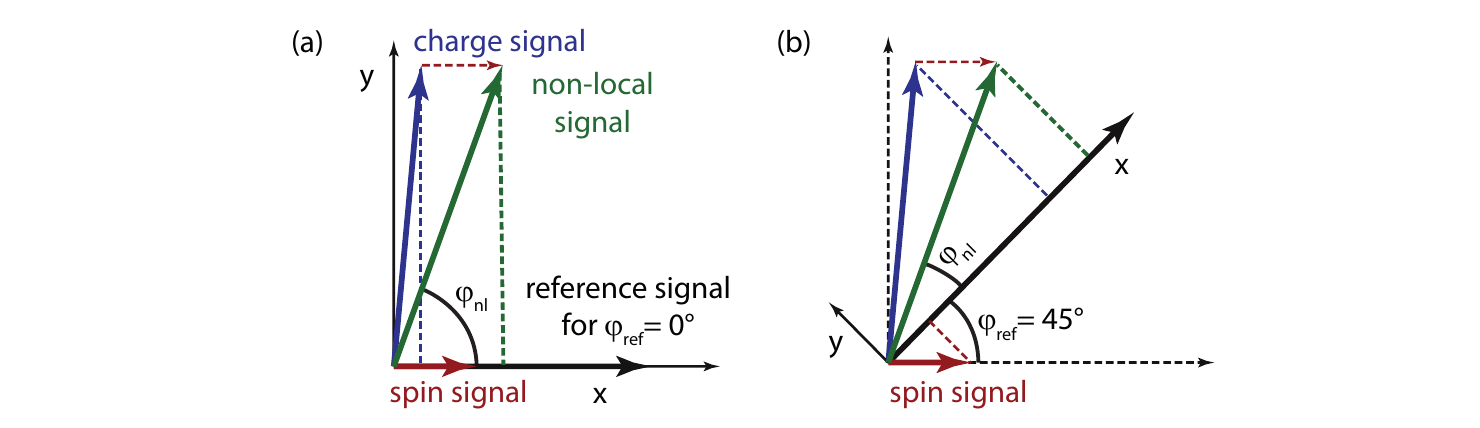}
	\caption{(Color online) Vector diagrams illustrating the components of the non-local voltage signal. (a) If the reference phase of the lock-in is set to the default value of $\varphi_{\text{ref}}=\unit{0}{\degree}$ (for further explanation see text), the spin signal is measured in the x-component of the lock-in. A phase shifted charge signal results in a total non-local voltage signal (sum of charge and spin signal), which is significantly phase shifted by $\varphi_{\text{nl}}$ relative to the x-channel. (b) By changing the reference phase $\varphi_{\text{ref}}$ of the lock-in amplifier, its x- and y-channels are rotated relative to both charge and spin signals. For $\varphi_{\text{ref}}=\unit{45}{\degree}$, this results in an increase of the charge signal and a decrease of the spin signal in the x-channel which are both projected on the new x-axis direction. }
	\label{fig:fig5}
\end{figure}

For our standard non-local spin measurement, we therefore record the signal in the x-channel of the lock-in amplifier at a reference phase of $\varphi_{\text{ref}}\approx 0$. This situation is illustrated in figure \ref{fig:fig5}(a). For $\varphi_{\text{ref}}\approx 0$ the reference signal is in phase with the current (see figure \ref{fig:fig4}(b)). Accordingly, the full spin signal is detected in the x-channel. Nevertheless, we usually measure a large phase $\varphi_{\text{nl}}$ of the overall non-local signal.

We attribute this out-of-phase signal to a pure charge-driven signal. It has a larger amplitude than the spin signal. Both charge and spin signals are sinusoidal and have the same frequency. Hence, their superposition is also sinusoidal as seen for the green curve in figure \ref{fig:fig4}(c).

Our assignment can be tested by changing the reference phase $\varphi_{\text{ref}}$, which will rotate the x-y-coordinate system relative to the phase of the internal oscillator which is in phase to the injection current. This is illustrated in figure \ref{fig:fig5}(b) for $\varphi_{\text{ref}}=\unit{45}{\degree}$. If the x-channel is still used to record the non-local voltage signal, its spin part will decrease and its charge contribution will increase, which can be seen by the respective projections of both contributions to the new x-axis direction which was rotated by $\varphi_{\text{ref}}=\unit{45}{\degree}$.

\begin{figure}[tb]
	\includegraphics{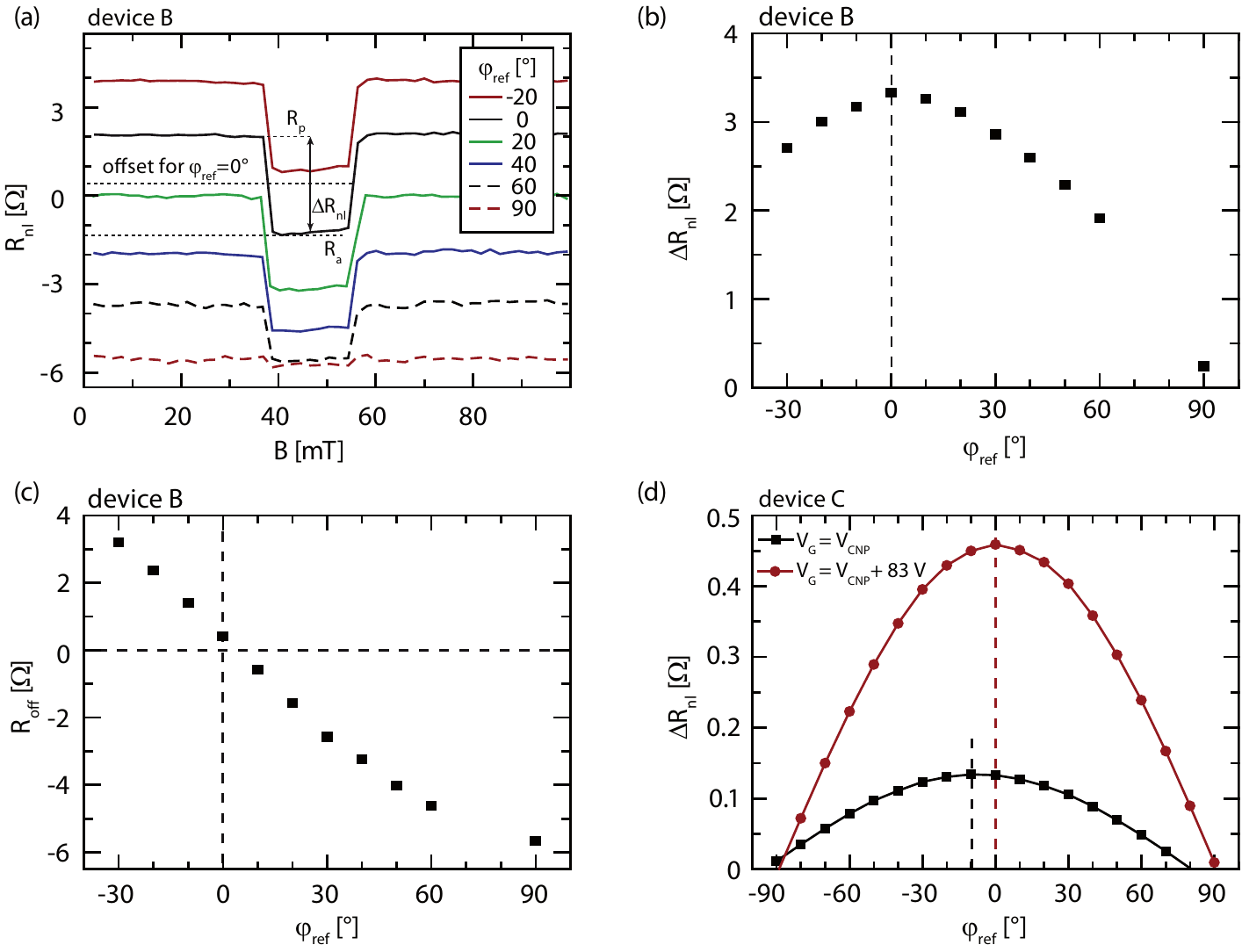}
	\caption{(Color online) (a) x-component of the lock-in amplifier during spin-valve measurements of device B at different reference phases (actual data set; no waterfall diagram). The amplitude of the spin signal decreases (b) whereas the background (sometimes also called basline resistance) increases (c) for reference phases away from $\varphi_{\text{ref}}=\unit{0}{\degree}$. This demonstrates that the non-local voltage is a composition of a spin signal and a phase shifted charge signal like depicted in figure \ref{fig:fig5}. (d) Amplitude of the spin signal as a function of the reference phase for device C at the charge neutrality point and at very high charge carrier densities.}
	\label{fig:fig6}
\end{figure}

In figure \ref{fig:fig6}(a) we show a series of non-local spin transport measurements where we changed the reference phase $\varphi_{\text{ref}}$ only. In contrast to the Hanle spin precession measurements discussed above, we now apply an in-plane magnetic field which is oriented parallel to the magnetic electrodes. All curves in figure \ref{fig:fig6}(a) are plotted as measured. No vertical offset was added. For $\varphi_{\text{ref}}=\unit{0}{\degree}$, the parallel component of the non-local signal is at $R_{\text{p}}=\unit{2.1}{\ohm}$ while the anti-parallel component is at $R_{\text{a}}=\unit{-1.3}{\ohm}$. Accordingly, the magnitude of the spin signal is $\Delta R_{\text{nl}}=R_{\text{p}}-R_{\text{a}}=\unit{3.3}{\ohm}$ with a background signal of $R_{\text{off}}=(R_{\text{p}}+R_{\text{a}})/2=\unit{0.4}{\ohm}$ (see dotted line in figure \ref{fig:fig6}(a)).

The spin signal $\Delta R_{\text{nl}}$ indeed decreases for both positive and negative $\varphi_{\text{ref}}$ as seen in figure \ref{fig:fig6}(b) while the offset baseline resistance $R_{\text{off}}$ changes with $\varphi_{\text{ref}}$ and switches sign at about $\varphi_{\text{ref}}=\unit{3}{\degree}$ (figure \ref{fig:fig6}(c)). This finding is in full accordance with the vector diagrams discussed in figure \ref{fig:fig5} and shows again that the phases of spin signal and the charge-driven background resistance differ strongly. In figure \ref{fig:fig6}(d) we show the amplitude of the spin signal as a function of the reference phase for another device taken at different back gate voltages, i.e. different carrier densities.

At large electron doping (red curve) the maximum of the spin signal is still at $\varphi_{\text{ref}}=\unit{0}{\degree}$ while it shifts towards $\varphi_{\text{ref}}=-\unit{10}{\degree}$ near the charge neutrality point (black curve). Currently, we do not have an explanation for this additional phase shift. We note, however, that not all devices show this shift. The one presented in figure \ref{fig:fig6}(d) is the largest we observed in our devices. Furthermore, the error on the magnitude of the measured spin signal is quite low (cos$(\unit{10}{\degree})\approx 0.985$) if the reference phase is kept fixed for back gate dependent spin transport measurements.

\begin{figure}[tb]
	\includegraphics{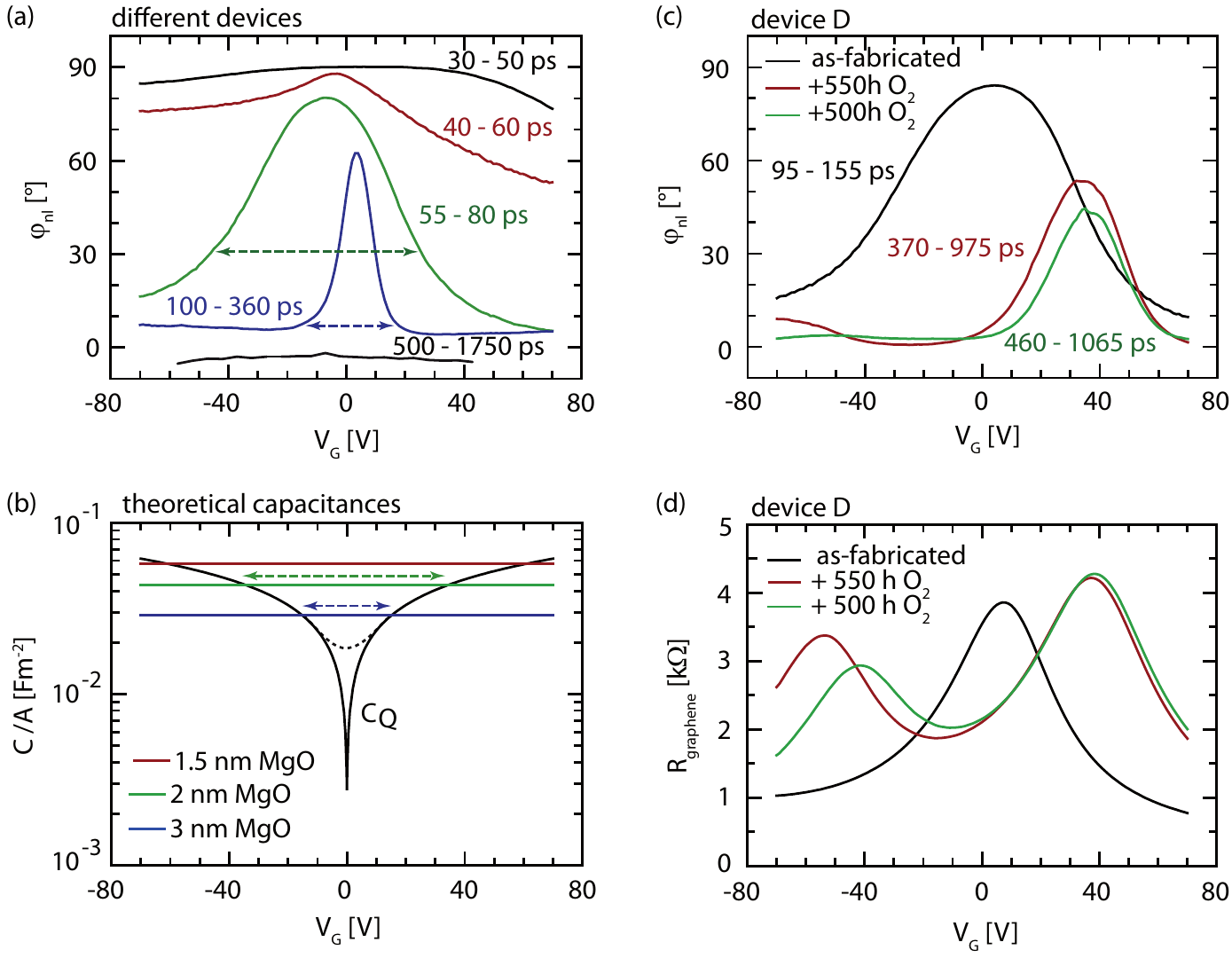}
	\caption{(Color online) (a) Gate voltage dependence of the phase of non-local voltage for different spin transport devices taken at room temperature.  The overall phase decreases for devices with longer spin lifetimes. The back gate dependent total change of the spin lifetime is given for each device. (b) Calculated oxide barrier capacitances for different oxide layer thicknesses and gate dependent quantum capacitance of graphene. The intersections between oxide barrier and graphene capacitance seem to determine the width of the phase curves. The dashed lines with arrows have the same length as in (a). (c) Non-local phase vs. back gate voltage for device D before and after oxygen treatments. Such a treatment improves the oxide barrier, which in turn leads to a significant increase in spin lifetime and a decrease in the measured phase (device is discussed in more detail in \cite{PhysRevB.90.165403}). (d) Graphene resistance vs. back gate voltage for the same device. The strong change in phase is only seen near the right CNP after each oxygen treatment (green and red curve).  }
	\label{fig:fig7}
\end{figure}

We next explore the influence of the Co/MgO-to-graphene interface quality on the non-local charge signal. Our previous spin transport studies indicated that the quality of the oxide barrier is determined by the magnitude of the contact resistance area product $R_cA$ as the measured spin lifetime $\tau_s$ increases with increasing $R_cA$ values \cite{PhysRevB.88.161405,PhysRevB.90.165403} (see also figure \ref{fig:fig1}(b)). We now focus on the charge density dependent phase $\varphi_{\text{nl}}$ of the non-local voltage signal, which is a measure of the charge-driven contribution. Figure \ref{fig:fig7}(a) shows the backgate dependent $\varphi_{\text{nl}}$ for a series of devices with increasing spin lifetimes, i.e. increasing $R_cA$ values. It is apparent that the overall phase becomes lower for devices with better oxide barrier quality, i.e. larger $R_cA$ and $\tau_s$ values. This is not only observed by an error prone comparison of different devices but can also be observed in single devices where the oxide barrier was improved by subsequent oxygen treatments (see figure \ref{fig:fig7}(c) and reference \cite{PhysRevB.90.165403}) which in turn yield longer spin lifetimes and larger $R_cA$ values.

Interestingly, devices with both high and low non-local phases ($\varphi_{\text{nl}}\approx\unit{0}{\degree}$ or $\unit{90}{\degree}$, figure \ref{fig:fig7}(a)) only exhibit a small gate dependence of the phase, whereas devices with intermediate phases show distinct gate dependent behavior. The maximum of $\varphi_{\text{nl}}$ is always around the charge neutrality point that can be attributed to the graphene part between the contacts. A contact-induced charge neutrality point (seen e.g. in the red and green curve in figure \ref{fig:fig7}(d) at large negative gate voltage; further details are found in reference \cite{PhysRevB.90.165403}) by far has not such an impact on the phase (see corresponding curves in figure \ref{fig:fig7}(c)). Combined with the observation that devices with overall high or low phases show almost no variation of the phase although they exhibit normal change in resistance as a function of gate voltage, the increase in phase cannot be explained by just an increase in device resistance.

The dependence of the non-local phase on both the oxide barrier quality and the gate voltage may imply that $\varphi_{\text{nl}}$ is given by the interplay between the capacitance of the MgO layer and the quantum capacitance of the graphene sheet. The quantum capacitance of an ideal, undoped graphene sheet \cite{luryi:501,fang:092109} is shown in figure \ref{fig:fig7}(b). Also depicted is the capacitance of MgO oxide barriers calculated by the plate capacitor model with the permittivity of bulk MgO and different MgO layer thicknesses (\unit{1.5-3}{\nano\meter}).
In contrast to the capacitance of the oxide barriers the quantum capacitance of graphene  depends on the Fermi energy which is tuned by the back gate voltage.

The intersections between the capacitance lines of the oxide barrier and graphene capacitance may determine the width of the non-local phase curves in figure \ref{fig:fig7}(a). Note that corresponding curves are plotted in the same color and dashed lines with arrows have the same length. This behavior can qualitatively be understood when treating the oxide barrier and the graphene as two capacitances put in series. In such a case, the smaller one dominates the total capacitance ($1/C_{\text{tot}}=1/C_1+1/C_2$). Accordingly, the quantum capacitance of the graphene sheet in between the contacts dominates near the CNP which might be the origin for the strong back gate dependence of $\varphi_{\text{nl}}$. We note that the small quantum capacitance around $V_{\text{G}}=\unit{0}{\volt}$ in figure \ref{fig:fig7}(b) is only a result of the vanishing charge carrier density of ideal graphene at the CNP. Experimental measurements of the quantum capacitance at room temperature show a broadening around the CNP \cite{Xu2011,Xia2009a}, which is illustrated by the dashed line in figure \ref{fig:fig7}(b). It is therefore suggestive that the oxide capacitance of the device with the largest spin lifetimes and lowest overall phase (see black curve in \ref{fig:fig7}(a)) is comparable or even smaller than the graphene quantum capacitance at all gate voltages. Therefore, this device does not show any significant back gate dependence of $\varphi_{\text{nl}}$. Furthermore, we note that the data of this device were measured almost two years after device fabrication. During that time, the device was stored both in vacuum and in dry air. In reference \cite{PhysRevB.90.165403} we described in detail the significant improvement in the spin transport properties by this long term oxidation. It is indicative that this long term oxidation have lead to very homogenous oxide barriers which in turn may be responsible for the low non-local phase and large spin lifetimes.

\section{Conclusions}

We demonstrated that non-local spin measurements in graphene spin valve devices are by far not free from charge effects as sometimes claimed in literature. On the one hand, these charge effects can result in unwanted signals in putative spin transport signals, i.e. they create a background in Hanle spin precession curves or they can lead to an underestimation of the spin signal if the lock-in amplifier is erroneously put to the phase of the charge signal.

On the other hand, these charge effects may also serve as a fingerprint to evaluate the quality of the oxide barriers. This is very important because oxide barriers in spin-valve structures are known to be a possible bottleneck for spin transport. The characterization of spin injection and spin detection barriers of a fabricated device is not a trivial matter, in particular if the existence of conducting pinholes within the barrier should be evaluated. Concerning this issue, we demonstrated that a part of the charge-driven signal is caused by the combination of inhomogeneous charge injection through conducting pinholes over the Co/MgO-to-graphene interface and the subsequent redistribution of the currents by a magnetic field similar to the classical Hall effect. Furthermore, we conducted a systematic study on the phase of the non-local spin and charge voltage when utilizing ac lock-in techniques and showed that it is also linked to the quality of the oxide barrier.
In summary, analyzing both the charge-induced background signals and the phase in non-local spin measurements provide important additional information about the quality of the oxide barrier in graphene-based non-local spin-valve devices which may help to unveil complicated spin transport phenomena.

\ack

The research leading to these results has received funding from the DFG through FOR-912 and the European Union Seventh Framework Programme under grant agreement n${^\circ}$604391 Graphene Flagship.

\section*{References}

\bibliography{Literatur}

\begin{thebibliography}{10}

\bibitem{PhysRevLett.105.167202}
Wei Han, K.~Pi, K.~M. McCreary, Yan Li, Jared J.~I. Wong, A.~G. Swartz, and
  R.~K. Kawakami.
\newblock Tunneling spin injection into single layer graphene.
\newblock {\em Phys. Rev. Lett.}, 105(16):167202, Oct 2010.

\bibitem{PhysRevB.86.235408}
T.~Maassen, I.~J. Vera-Marun, M.~H.~D. Guimar\~aes, and B.~J. van Wees.
\newblock Contact-induced spin relaxation in hanle spin precession
  measurements.
\newblock {\em Phys. Rev. B}, 86:235408, Dec 2012.

\bibitem{Dlubak2012}
Bruno Dlubak, Marie-Blandine Martin, Cyrile Deranlot, Bernard Servet, Stephane
  Xavier, Richard Mattana, Mike Sprinkle, Claire Berger, Walt~A. De~Heer,
  Frederic Petroff, Abdelmadjid Anane, Pierre Seneor, and Albert Fert.
\newblock Highly efficient spin transport in epitaxial graphene on sic.
\newblock {\em Nat Phys}, 8(7):557--561, July 2012.

\bibitem{PhysRevB.88.161405}
F.~Volmer, M.~Dr\"ogeler, E.~Maynicke, N.~von~den Driesch, M.~L. Boschen,
  G.~G\"untherodt, and B.~Beschoten.
\newblock Role of mgo barriers for spin and charge transport in co/mgo/graphene
  nonlocal spin-valve devices.
\newblock {\em Phys. Rev. B}, 88:161405, Oct 2013.

\bibitem{PhysRevB.90.165403}
F.~Volmer, M.~Dr\"ogeler, E.~Maynicke, N.~von~den Driesch, M.~L. Boschen,
  G.~G\"untherodt, C.~Stampfer, and B.~Beschoten.
\newblock Suppression of contact-induced spin dephasing in graphene/mgo/co
  spin-valve devices by successive oxygen treatments.
\newblock {\em Phys. Rev. B}, 90:165403, Oct 2014.

\bibitem{Akerman2002}
Johan~J. Akerman, R.~Escudero, C.~Leighton, S.~Kim, D.A. Rabson, Renu~Whig
  Dave, J.M. Slaughter, and Ivan~K. Schuller.
\newblock Criteria for ferromagnetic-insulator-ferromagnetic tunneling.
\newblock {\em Journal of Magnetism and Magnetic Materials}, 240(1-3):86--91,
  February 2002.

\bibitem{jonsson-akerman:1870}
B.~J. Jonsson-Akerman, R.~Escudero, C.~Leighton, S.~Kim, Ivan~K. Schuller, and
  D.~A. Rabson.
\newblock Reliability of normal-state current--voltage characteristics as an
  indicator of tunnel-junction barrier quality.
\newblock {\em Applied Physics Letters}, 77(12):1870--1872, 2000.

\bibitem{Rabson2001}
D.~A. Rabson, B.~J. Jonsson-Akerman, A.~H. Romero, R.~Escudero, C.~Leighton,
  S.~Kim, and Ivan~K. Schuller.
\newblock Pinholes may mimic tunneling.
\newblock {\em J. Appl. Phys.}, 89(5):2786--2790, March 2001.

\bibitem{Oliver2003}
Bryan Oliver, Qing He, Xuefei Tang, and Janusz Nowak.
\newblock Tunneling criteria and breakdown for low resistive magnetic tunnel
  junctions.
\newblock {\em J. Appl. Phys.}, 94(3):1783--1786, August 2003.

\bibitem{Droegeler2014}
Marc Dr\"ogeler, Frank Volmer, Maik Wolter, Bernat Terres, Kenji Watanabe,
  Takashi Taniguchi, Gernot G\"untherodt, Christoph Stampfer, and Bernd
  Beschoten.
\newblock Nanosecond spin lifetimes in single- and few-layer graphene-hbn
  heterostructures at room temperature.
\newblock {\em Nano Lett.}, 14(11):6050--6055, September 2014.

\bibitem{PhysRevLett.105.136601}
F.~L. Bakker, A.~Slachter, J.-P. Adam, and B.~J. van Wees.
\newblock Interplay of peltier and seebeck effects in nanoscale nonlocal spin
  valves.
\newblock {\em Phys. Rev. Lett.}, 105(13):136601, Sep 2010.

\bibitem{PhysRevB.76.153107}
Mark Johnson and R.~H. Silsbee.
\newblock Calculation of nonlocal baseline resistance in a
  quasi-one-dimensional wire.
\newblock {\em Phys. Rev. B}, 76(15):153107, Oct 2007.

\bibitem{PhysRevLett.101.046601}
N.~Tombros, S.~Tanabe, A.~Veligura, C.~Jozsa, M.~Popinciuc, H.~T. Jonkman, and
  B.~J. van Wees.
\newblock Anisotropic spin relaxation in graphene.
\newblock {\em Phys. Rev. Lett.}, 101(4):046601, Jul 2008.

\bibitem{doi:10.1021/nl301050a}
Marcos H.~D. Guimaraes, A.~Veligura, P.~J. Zomer, T.~Maassen, I.~J. Vera-Marun,
  N.~Tombros, and B.~J. van Wees.
\newblock Spin transport in high-quality suspended graphene devices.
\newblock {\em Nano Letters}, 12(7):3512--3517, 2012.

\bibitem{PhysRevB.80.214427}
M.~Popinciuc, C.~J\'ozsa, P.~J. Zomer, N.~Tombros, A.~Veligura, H.~T. Jonkman,
  and B.~J. van Wees.
\newblock Electronic spin transport in graphene field-effect transistors.
\newblock {\em Phys. Rev. B}, 80(21):214427, Dec 2009.

\bibitem{PhysRevB.87.081402}
M.~Wojtaszek, I.~J. Vera-Marun, T.~Maassen, and B.~J. van Wees.
\newblock Enhancement of spin relaxation time in hydrogenated graphene
  spin-valve devices.
\newblock {\em Phys. Rev. B}, 87:081402, Feb 2013.

\bibitem{PhysRevB.87.075455}
Adrian~G. Swartz, Jen-Ru Chen, Kathleen~M. McCreary, Patrick~M. Odenthal, Wei
  Han, and Roland~K. Kawakami.
\newblock Effect of \textit{in situ} deposition of mg adatoms on spin
  relaxation in graphene.
\newblock {\em Phys. Rev. B}, 87:075455, Feb 2013.

\bibitem{PhysRevLett.104.187201}
K.~Pi, Wei Han, K.~M. McCreary, A.~G. Swartz, Yan Li, and R.~K. Kawakami.
\newblock Manipulation of spin transport in graphene by surface chemical
  doping.
\newblock {\em Phys. Rev. Lett.}, 104(18):187201, May 2010.

\bibitem{han222109}
Wei Han, K.~Pi, W.~Bao, K.~M. McCreary, Yan Li, W.~H. Wang, C.~N. Lau, and
  R.~K. Kawakami.
\newblock Electrical detection of spin precession in single layer graphene spin
  valves with transparent contacts.
\newblock {\em Applied Physics Letters}, 94(22):222109, 2009.

\bibitem{PhysRevLett.109.186604}
Kathleen~M. McCreary, Adrian~G. Swartz, Wei Han, Jaroslav Fabian, and Roland~K.
  Kawakami.
\newblock Magnetic moment formation in graphene detected by scattering of pure
  spin currents.
\newblock {\em Phys. Rev. Lett.}, 109:186604, Nov 2012.

\bibitem{PhysRevLett.113.086602}
M.~H.~D. Guimar\~aes, P.~J. Zomer, J.~Ingla-Ayn\'es, J.~C. Brant, N.~Tombros,
  and B.~J. van Wees.
\newblock Controlling spin relaxation in hexagonal bn-encapsulated graphene
  with a transverse electric field.
\newblock {\em Phys. Rev. Lett.}, 113:086602, Aug 2014.

\bibitem{Idzuchi2012}
Hiroshi Idzuchi, Yasuhiro Fukuma, and YoshiChika Otani.
\newblock Towards coherent spin precession in pure-spin current.
\newblock {\em Sci. Rep.}, 2:--, September 2012.

\bibitem{ISI000249789600001}
Jaroslav Fabian, Alex Matos-Abiague, Christian Ertler, Peter Stano, and Igor
  \v{Z}uti\'{c}.
\newblock Semiconductor spintronics.
\newblock {\em ACTA PHYSICA SLOVACA}, 57({4-5}):565--907, {AUG-OCT} 2007.

\bibitem{PhysRevB.37.5312}
Mark Johnson and R.~H. Silsbee.
\newblock Coupling of electronic charge and spin at a
  ferromagnetic-paramagnetic metal interface.
\newblock {\em Phys. Rev. B}, 37(10):5312--5325, Apr 1988.

\bibitem{doi:10.1021/nl203733r}
Yung-Chang Lin, Chun-Chieh Lu, Chao-Huei Yeh, Chuanhong Jin, Kazu Suenaga, and
  Po-Wen Chiu.
\newblock Graphene annealing: How clean can it be?
\newblock {\em Nano Letters}, 12(1):414--419, 2012.

\bibitem{doi:10.1021/jz201653g}
Recep Zan, Ursel Bangert, Quentin Ramasse, and Konstantin~S. Novoselov.
\newblock Interaction of metals with suspended graphene observed by
  transmission electron microscopy.
\newblock {\em The Journal of Physical Chemistry Letters}, 3(7):953--958, 2012.

\bibitem{PhysRevB.79.245430}
T.~Mueller, F.~Xia, M.~Freitag, J.~Tsang, and Ph. Avouris.
\newblock Role of contacts in graphene transistors: A scanning photocurrent
  study.
\newblock {\em Phys. Rev. B}, 79(24):245430, Jun 2009.

\bibitem{luryi:501}
Serge Luryi.
\newblock Quantum capacitance devices.
\newblock {\em Applied Physics Letters}, 52(6):501--503, 1988.

\bibitem{fang:092109}
Tian Fang, Aniruddha Konar, Huili Xing, and Debdeep Jena.
\newblock Carrier statistics and quantum capacitance of graphene sheets and
  ribbons.
\newblock {\em Applied Physics Letters}, 91(9):092109, 2007.

\bibitem{Xu2011}
Huilong Xu, Zhiyong Zhang, Zhenxing Wang, Sheng Wang, Xuelei Liang, and
  Lian-Mao Peng.
\newblock Quantum capacitance limited vertical scaling of graphene field-effect
  transistor.
\newblock {\em ACS Nano}, 5(3):2340--2347, February 2011.

\bibitem{Xia2009a}
Fengnian Xia, Thomas Mueller, Yu-ming Lin, Alberto Valdes-Garcia, and Phaedon
  Avouris.
\newblock Ultrafast graphene photodetector.
\newblock {\em Nat Nano}, 4(12):839--843, December 2009.

\end{thebibliography}
\bibliographystyle{unsrt}

\end{document}